\documentclass{article}
\usepackage{amssymb}
\usepackage{amsmath}

\input{tcilatex}

\begin{document}

\title{On the ADHM construction of noncommutative $U(2)$ k-instanton}
\date{}
\author{M.Lagraa\thanks{%
m.lagraa@mailcity.lycos.com} \\
Laboratoire de physique theorique\\
Universit\'{e} d'Oran Es-Senia, 31100, Algerie.}
\maketitle

\begin{abstract}
The basic objects of the ADHM construction are reformulated in terms of
elements of the $\mathcal{A}_{\theta }(\mathbb{R}^{4})$ algebra of the
noncommutative $\mathbb{R}_{\theta }^{4}$ space. This new formulation of the
ADHM construction makes possible the explicit calculus of the $U(2)$
instanton number which is shown to be the product of a trace of a finite
rank projector of the Fock representation space of the algebra $\mathcal{A}%
_{\theta }(\mathbb{R}^{4})$ times a noncommutative version of the winding
number.

\textbf{PACS\ NUMBER:} 11.10.Nx, 11.15.Tk.
\end{abstract}

\begin{quotation}
\textbf{Introduction }
\end{quotation}

The ADHM (Atiyah-Drinfeld-Hitchin-Manin)\ construction of instantons \cite%
{ADHM} is an active research area in physics. It emerges from a wide range
of phenomenon in physics, from the Dp-brane theory where the space of
solutions of the Higgs branch of the Dp-brane coincides with the ADHM
equations \cite{Witten}\cite{Douglas1}\cite{Douglas2} to the vortex theory
where it is shown that it exists relationship between the moduli space of
vortices and the moduli space of instantons given in terms of fields of the
ADHM construction \cite{Hanani1}\cite{hanani2}\cite{Eto}.

In recent years many works have been devoted to the calculus of the
noncommutative ADHM instanton number. Either from Corrigan'identity \cite%
{Corrigan1}\cite{CHU}\cite{TIAN} or in the operator formalism where the
first Pontrjagin class is calculated as a converge series \cite{furuuchi1}%
\cite{furuuchi3}\cite{furuuchi4}\cite{Sako1}\cite{Sako2}. In this work we
reformulate the ADHM construction of instantons in terms of elements of the $%
\mathcal{A}_{\theta }(\mathbb{R}^{4})\otimes \mathcal{A}_{\theta }(\mathbb{R}%
^{4})$ algebra. This new formulation leads to explicit solutions of ADHM
constraints. It makes possible an analytic calculus of the noncommutative $%
U(2)$ gauge field, the field strength and the $U(2)$ instanton number. The
calculus of this instanton number resembles the one of the element of the
third homotopy group $\pi _{3}(SU(2))$: the winding number measured by the
noncommutative version of the three dimensional surface integral at
infinity. This suggests a geometrical picture of instanton number of the
ADHM construction which is shown to be a product of a trace of a finite rank
projector onto the Fock representation space of the algebra $\mathcal{A}%
_{\theta }(\mathbb{R}^{4})$ times a noncommutative version of the winding
number.

We begin this paper by recalling, in section 1, some properties of the
noncommutative $\mathbb{R}_{\theta }^{4}$ space and review briefly, in
section 2, the\ ADHM construction \ of noncommutative instantons\ \cite%
{nekrasov0}. In section \ 3, we reformulate the noncommutative ADHM
construction of instantons in terms of elements of the $\mathcal{A}_{\theta
}(\mathbb{R}^{4})\otimes \mathcal{A}_{\theta }(\mathbb{R}^{4})$ algebra from
which we calculate analytically the $U(2)$ gauge field, the field strength
and the $U(2)$ instanton number then we show how it may be described by a
noncommutative version of the winding number.

\section{Noncommutative $\mathbb{R}_{\protect\theta }^{4}$}

The noncommutative 4-dimensional $\mathbb{R}_{\theta }^{4}$ space is
described by the $\mathcal{A}_{\theta }(\mathbb{R}^{4})$ algebra generated
by the coordinates (hermitian operators) $x_{\mu }(\mu \in \left[ 1,2,3,4%
\right] )$ or with the complex notation by

\begin{equation}
z_{1}=x_{2}+ix_{1},\text{ \ }z_{2}=x_{4}+ix_{3},\text{ \ }\overline{z}%
_{1}=x_{2}-ix_{1},\text{ \ }\overline{z}_{2}=x_{4}-ix_{3}
\label{zcordinates}
\end{equation}%
with the commutations rules 
\begin{equation}
\left[ z_{\alpha },\overline{z}_{\beta }\right] =-2\theta \delta _{\alpha
,\beta },\text{ }\left[ z_{\alpha },z_{\beta }\right] =0\text{ \ },(\alpha
=1,2).  \label{zcomm}
\end{equation}

The derivatives act on the algebra $\mathcal{A}_{\theta }(\mathbb{R}^{4})$ as%
\begin{equation}
\partial _{\alpha }a=\frac{1}{2\theta }\left[ \overline{z}_{\alpha },a\right]
\text{, \ \ \ \ \ \ }\partial _{\overline{\alpha }}=\frac{-1}{2\theta }\left[
z_{\alpha },a\right] \text{ \ \ }\forall a\in \mathcal{A}_{\theta }(\mathbb{R%
}^{4})
\end{equation}%
from which we define the action of exterior derivative $d$ by

\begin{equation*}
da=\frac{1}{2\theta }\left[ \overline{z}_{\alpha },a\right] dz_{\alpha }-%
\frac{1}{2\theta }\left[ z_{\alpha },a\right] d\overline{z}_{\alpha }.
\end{equation*}

$dz_{\alpha }$ and $d\overline{z}_{\alpha }$ commute with $z_{\alpha }$ and $%
\overline{z}_{\alpha }$ and anti-commute among themselves, and hence $%
d^{2}a=0$ $\forall a\in \mathcal{A}_{\theta }(\mathbb{R}^{4}).$

The Fock representation space $\mathcal{H}$ of the algebra $\mathcal{A}%
_{\theta }(\mathbb{R}^{4})$ is spanned by the orthonormalized basis $\
\left\vert n_{1},n_{2}\right\rangle $ $\left( n_{1}\geq 0,n_{2}\geq 0\right)
,$ $\langle n_{1},n_{2}\left\vert m_{1},m_{2}\right\rangle =\delta
_{n_{1}m_{1}}\delta _{n_{2}m_{2}}$, with

\begin{eqnarray}
z_{1}\left\vert n_{1},n_{2}\right\rangle &=&\sqrt{2\theta \left(
n_{1}+1\right) }\left\vert n_{1}+1,n_{2}\right\rangle ,\text{ }%
z_{2}\left\vert n_{1},n_{2}\right\rangle =\sqrt{2\theta \left(
n_{2}+1\right) }\left\vert n_{1},n_{2}+1\right\rangle ,  \notag \\
\overline{z}_{1}\left\vert n_{1},n_{2}\right\rangle &=&\sqrt{2\theta n_{1}}%
\left\vert n_{1}-1,n_{2}\right\rangle ,\text{ }\overline{z}_{2}\left\vert
n_{1},n_{2}\right\rangle =\sqrt{2\theta n_{2}}\left\vert
n_{1},n_{2}-1\right\rangle .  \label{REPN}
\end{eqnarray}

In the complex notation the integration on $\mathbb{R}_{\theta }^{4}$ is
defined as:

\begin{equation}
\int dz_{1}d\overline{z}_{1}dz_{2}d\overline{z}_{2}=(4\pi \theta )^{2}Tr_{%
\mathcal{H}}  \label{integ defi}
\end{equation}%
where the trace of the operator is over the Fock space $\mathcal{H}$.

\section{Instantons and\ \textbf{ADHM} construction}

Instantons are localized finite-action non-perturbative self(anti-self)-dual
solutions for the Euclidian equations of motion of Yang-Mills gauge
theories. In this section we will recall the basic algoritm of the ADHM
construction \cite{nekrasov0} to give such solutions on noncommutative $%
\mathbb{R}_{\theta }^{4}$ space. This construction is just a deformed
version of the commutative one \cite{ADHM}\cite{Corrigan}.

The different steps of the ADHM construction for $U(N)$ $k$-instantons can
be summarized as follows

\begin{enumerate}
\item Solve the deformed ADHM equations 
\begin{equation}
\begin{array}{l}
\text{ \ \ }\left[ B_{1},B_{1}^{\dagger }\right] +\left[ B_{2},B_{2}^{%
\dagger }\right] +II^{\dagger }-J^{\dagger }J=4\theta id_{k}. \\ 
\text{ \ \ }\left[ B_{1},B_{2}\right] +IJ=0.%
\end{array}
\label{ADHMB}
\end{equation}%
where we have the matrices (with $\mathcal{C}$-numbers entries) 
\begin{eqnarray*}
B_{1,2} &:&k\times k\text{ dimension.} \\
I,J^{\dagger } &:&k\times N\text{ dimension.} \\
id_{k} &:&\text{ the }k\times k\text{ identity.}
\end{eqnarray*}

\item Define the Dirac-like operator 
\begin{equation}
\mathcal{D}_{z}=\left( 
\begin{array}{ccc}
B_{2}-z_{2} & B_{1}-z_{1} & I \\ 
-(B_{1}^{\dagger }-\overline{z}_{1}) & B_{2}^{\dagger }-\overline{z}_{2} & 
J^{\dagger }%
\end{array}%
\right) .  \label{DIRACB}
\end{equation}

\item Look for all the $N$ normalized solutions $\Psi ^{a}$ (the zero-modes)
to the equation 
\begin{equation}
\mathcal{D}_{z}\Psi ^{a}=0,\text{ \ \ }\Psi ^{a\dagger }\Psi
^{b}=id_{k}\delta ^{ab}.  \label{NORMPSIADHM}
\end{equation}

\item Construct the $U(N)$ gauge field $A=\Psi ^{\dagger }d\Psi $ from which
we define the field strength $F=dA+A^{2}$.
\end{enumerate}

The instanton number is defined by

\begin{equation}
k=\frac{\pm 1}{8\pi ^{2}}\left( 4\pi \theta \right) ^{2}Tr\left( F\right)
^{2}  \label{kINST}
\end{equation}%
which takes integral value. Here the sign +(-) is for the (anti-)self-dual
instantons and the trace is taken both on the group indices (for the general
case of the $U(N)$ gauge group) and on the Fock space. The above relation is
the noncommutative version of the second Chern character defined by

\begin{equation*}
k=\frac{\pm 1}{8\pi ^{2}}\int_{R^{4}}dxTr_{U(N)}\left( F\right) ^{2}
\end{equation*}

\section{$U(2)$-k-Instanton explicit solution}

In general, in the commutative case ($\theta =0)$, we solve the ADHM
equation (\ref{ADHMB}) by putting the $k\times k$ matrices $B_{1}$ and $%
B_{2} $ diagonal. Their $k$ complex eingenvalues $\alpha
_{1}^{i}=X_{2}^{i}+iX_{1}^{i}$, $\alpha _{2}^{i}=X_{4}^{i}+iX_{3}^{i}$ $%
(i\in \left[ 1,...,k\right] )$ are interpreted as positions of $k$
instantons (see \cite{HAMANAKA}). The elements of the matrices $I$ and $J$
give their size. Then it might be tempted to interpret, in the
noncommutative case, $B_{1}$ and $B_{2}$ as positions in the noncommutative $%
\mathbb{R}_{\theta }^{4}$ space i.e. $B_{1}$, $B_{2}\in $ $\mathcal{A}%
_{\theta }(\mathbb{R}^{4})$. The finite dimensions of the ADHM matrices are
obtained by using projectors of finite rank. In fact if we consider $%
\mathcal{A}_{\theta }(\mathbb{R}^{4})$ algebra elements of the form $%
z_{\alpha }^{P}=Pz_{\alpha }P$ where $P$ is the projector of finite rank

\begin{equation}
P=\sum_{n_{1}=N_{1}}^{N_{2}}\sum_{n_{2}=M_{1}}^{M_{2}}\left\vert
n_{1},n_{2}\right\rangle \left\langle n_{1},n_{2}\right\vert ,  \label{PROJ}
\end{equation}%
we obtain the following commutation relations

\begin{equation}
\lbrack z_{\alpha }^{P},\overline{z}_{\beta }^{P}]=-2\theta P\delta _{\alpha
\beta }+(I_{\alpha }I_{\alpha }^{\dagger }-J_{\alpha }^{\dagger }J_{\alpha
})\delta _{\alpha \beta },\text{ \ \ \ \ \ \ \ }[z_{\alpha }^{P},z_{\beta
}^{P}]=0.  \label{comzp}
\end{equation}

These relations follow from

\begin{equation}
z_{\alpha }P=Pz_{\alpha }+I_{\alpha }^{\dagger }-J_{\alpha }^{\dagger }\text{
\ \ , \ \ }\overline{z}_{\alpha }P=P\overline{z}_{\alpha }-I_{\alpha
}+J_{\alpha }  \label{zppz}
\end{equation}%
obtained from (\ref{REPN}) and (\ref{PROJ}) where

\begin{eqnarray}
I_{1} &=&P_{N_{2}}\overline{z}_{1}=\sum_{n_{2}=M_{1}}^{M_{2}}\left\vert
N_{2},n_{2}\right\rangle \left\langle N_{2},n_{2}\right\vert \text{ }%
\overline{z}_{1}\text{\ ,}  \notag \\
I_{2} &=&P_{M_{2}}\overline{z}_{2}=\sum_{n_{1}=N_{1}}^{N_{2}}\left\vert
n_{1},M_{2}\right\rangle \left\langle n_{1},M_{2}\right\vert \text{ }%
\overline{z}_{2},  \notag \\
J_{1} &=&\overline{z}_{1}P_{N_{1}}=\sum_{n_{2}=M_{1}}^{M_{2}}\overline{z}%
_{1}\left\vert N_{1},n_{2}\right\rangle \left\langle N_{1},n_{2}\right\vert 
\text{,}  \notag \\
J_{2} &=&\overline{z}_{2}P_{M_{1}}=\sum_{n_{1}=N_{1}}^{N_{2}}\overline{z}%
_{2}\left\vert n_{1},M_{1}\right\rangle \left\langle n_{1},M_{1}\right\vert 
\text{.}  \label{I1J1ADHM}
\end{eqnarray}

The operators $I_{\alpha }$ and $J_{\alpha }$ satisfy the relations

\begin{equation}
I_{\alpha }J_{\beta }=0,  \label{IJ0}
\end{equation}

and

\begin{eqnarray}
I_{1}I_{1}^{\dagger } &=&2\theta (N_{2}+1)P_{N_{2}},\text{ }%
I_{2}I_{2}^{\dagger }=2\theta (M_{2}+1)P_{M_{2}},  \notag \\
J_{1}^{\dagger }J_{1} &=&2\theta N_{1}P_{N_{1}},\text{ }J_{2}^{\dagger
}J_{2}=2\theta M_{1}P_{M_{1}}.  \label{IICROI}
\end{eqnarray}

Comparing equations (\ref{ADHMB}), (\ref{comzp}) and (\ref{IJ0}) above, we
see that we can identify $B_{\alpha }\in End(%
\mathbb{C}
^{k})$ with $\overline{z}_{\alpha }^{P}$, the Fock sub-space $P\mathcal{H}$
with the $k=(N_{2}-N_{1}+1)(M_{2}-M_{1}+1)-$dimensional space $%
\mathbb{C}
^{k}=V$ and the projector $P=id_{P\mathcal{H}}$ with $id_{k}$ to get the
solutions to the ADHM constraints under the form:

\begin{eqnarray}
\lbrack \overline{z}_{1}^{P},z_{1}^{P}]+[\overline{z}%
_{2}^{P},z_{2}^{P}]+(II^{\dagger }-J^{\dagger }J) &=&4\theta P=4\theta id_{k}
\notag \\
\lbrack z_{1}^{P},z_{2}^{P}]+IJ &=&0  \label{ADHMz}
\end{eqnarray}%
where $II^{\dagger }=I_{1}I_{1}^{\dagger }+I_{2}I_{2}^{\dagger }$,$\
J^{\dagger }J=J_{1}^{\dagger }J_{1}+J_{2}^{\dagger }J_{2}$ and $%
IJ=I_{1}J_{1}+I_{2}J_{2}$. The space $W$ is identified to

\begin{equation*}
W=((z_{1}\overline{z}_{1})^{-1}z_{1}P_{N_{2}}+(z_{2}\overline{z}%
_{2})^{-1}z_{2}P_{M_{2}}+(z_{1}\overline{z}_{1}+2\theta )^{-1}\overline{z}%
_{1}P_{N_{1}}+(z_{2}\overline{z}_{2}+2\theta )^{-1}\overline{z}_{2}P_{M_{1}})%
\mathcal{H}
\end{equation*}%
then $I_{\alpha }\in Hom(W,V)$, and $J_{\alpha }^{\dagger }\in Hom(W,V)$ as
required by ADHM construction. With this identification, the Dirac operator (%
\ref{DIRACB}) becomes 
\begin{equation}
\mathcal{D}_{z}=\left( 
\begin{array}{c}
\begin{array}{cccc}
Z_{2} & Z_{1} & I_{1} & I_{2}%
\end{array}
\\ 
\begin{array}{cccc}
-\overline{Z}_{1} & \overline{Z}_{2} & J_{1}^{\dagger } & J_{2}^{\dagger }%
\end{array}%
\end{array}%
\right)  \label{DIRACZ}
\end{equation}%
where $Z_{\alpha }=\overline{z}_{\alpha }^{P}\otimes id_{\mathcal{H}}-id_{P%
\mathcal{H}}\otimes z_{\alpha }$ , $\overline{Z}_{\alpha }=z_{\alpha
}^{P}\otimes id_{\mathcal{H}}-id_{P\mathcal{H}}\otimes \overline{z}_{\alpha
},$ $I_{\alpha }=I_{\alpha }\otimes id_{\mathcal{H}}$, $J_{\alpha
}=J_{\alpha }\otimes id_{\mathcal{H}}$ $\in $ $\mathcal{A}_{\theta }(\mathbb{%
R}^{4})\otimes \mathcal{A}_{\theta }(\mathbb{R}^{4})$.

The commutation rules (\ref{zcomm}) and (\ref{comzp}) can be recasted in

\begin{equation}
\lbrack Z_{\alpha },\overline{Z}_{\beta }]=-(I_{\alpha }I_{\alpha }^{\dagger
}-J_{\alpha }^{\dagger }J_{\alpha })\delta _{\alpha \beta },\ \ \ \ \
[Z_{\alpha },Z_{\beta }]=0.  \label{COMZ}
\end{equation}

From (\ref{COMZ}) and (\ref{IJ0}) we get the following two solutions to the
equation $\mathcal{D}_{z}\Psi =0$ as%
\begin{equation}
\psi ^{1}=\left( 
\begin{array}{c}
-\overline{Z}_{2}(Z\overline{Z})^{-1}II^{\dagger } \\ 
-\overline{Z}_{1}(Z\overline{Z})^{-1}II^{\dagger } \\ 
I_{1}^{\dagger } \\ 
I_{2}^{\dagger }%
\end{array}%
\right) \chi ^{-1}\text{ and \ }\psi ^{2}=\left( 
\begin{array}{c}
Z_{1}(\overline{Z}Z)^{-1}J^{\dagger }J \\ 
-Z_{2}(\overline{Z}Z)^{-1}J^{\dagger }J \\ 
J_{1} \\ 
J_{2}%
\end{array}%
\right) \phi ^{-1}  \label{SOLZ}
\end{equation}%
where $Z\overline{Z}=Z_{1}\overline{Z}_{1}+Z_{2}\overline{Z}_{2}$, $\chi
^{2}=II^{\dagger }(Z\overline{Z})^{-1}(Z\overline{Z}+II^{\dagger })$ and $%
\phi ^{2}=J^{\dagger }J(\overline{Z}Z)^{-1}(\overline{Z}Z+J^{\dagger }J)$
are the renormalization factors. The components of $\psi ^{1},\psi ^{2}\in $ 
$\mathcal{A}_{\theta }(\mathbb{R}^{4})\otimes \mathcal{A}_{\theta }(\mathbb{R%
}^{4})$ and act in the Fock space $P\mathcal{H\otimes H}$. By using (\ref%
{COMZ}) and (\ref{IJ0}) we can show that these solutions are orthogonal $%
\psi ^{\dagger 1}\psi ^{2}=0=\psi ^{\dagger 2}\psi ^{1}$. To satisfy the
normalization condition (\ref{NORMPSIADHM}) required by the ADHM
construction we must investigate the region of $P\mathcal{H\otimes H}$ where
the solutions (\ref{SOLZ}) are well-defined.

The solution $\psi ^{1}$ is well-defined on $range(II^{\dagger })=P^{\top }%
\mathcal{H\otimes H}\subset P\mathcal{H\otimes H}$ where $\ker (\overline{ZZ}%
)$ is projected out. The projector $P^{\top }$ is given

\begin{equation*}
P^{\top }=P_{N_{2}}+P_{M_{2}}-\left\vert N_{2},M_{2}\right\rangle
\left\langle N_{2},M_{2}\right\vert
\end{equation*}

The states belonging to $\ker (\overline{Z}_{\alpha })$ are coherent states
given by

\begin{equation*}
\left\vert C_{k,l}\right\rangle =\exp (\frac{z_{\alpha }^{P}}{\sqrt{2\theta }%
}\otimes \frac{z_{\alpha }}{\sqrt{2\theta }})\left\vert k,l\right\rangle
\otimes \left\vert 0,0\right\rangle ,\text{ \ }N_{1}\leq k\leq N_{2},\text{ }%
M_{1}\leq l\leq M_{2}.
\end{equation*}

Due to $(z_{1}^{P})^{N_{2}-k+1}\left\vert k,l\right\rangle =0$ and $%
(z_{2}^{P})^{M_{2}-l+1}\left\vert k,l\right\rangle =0$, the development of
the exponential is a finite sum leading to the following explicit forms of
the coherent states

\begin{equation*}
\left\vert C_{k,l}\right\rangle
=\sum_{n_{1}=0}^{N_{2}-k}\sum_{n_{2}=0}^{M_{2}-l}\left( \sqrt{\frac{%
(n_{1}+k)!}{n_{1}!k!}}\right) ^{\frac{1}{2}}\left( \sqrt{\frac{(n_{2}+l)!}{%
n_{2}!l!}}\right) ^{\frac{1}{2}}\left\vert n_{1}+k,n_{2}+l\right\rangle
\otimes \left\vert n_{1},n_{2}\right\rangle
\end{equation*}%
normalized as

\begin{equation*}
|\widetilde{C}_{k,l}\rangle =\left\vert C_{k,l}\right\rangle \left\langle
C_{k,l}\right\vert C_{k,l}\rangle ^{\frac{-1}{2}}\text{ \ \ , \ \ }\langle 
\widetilde{C}_{k,l}|\widetilde{C}_{k^{^{\prime }},l^{^{\prime }}}\rangle
=\delta _{k,k^{\prime }}\delta _{l,l^{\prime }}.
\end{equation*}

The sub-space $P^{C}\mathcal{H\otimes H}\subset P\mathcal{H\otimes H}$ plays
the roles of the vacuum for the algebra (\ref{COMZ}) where the projector

\begin{equation*}
P^{C}=\sum_{k=N_{1}}^{N_{2}}\sum_{l=M_{1}}^{M_{2}}\left\vert \widetilde{C}%
_{k,l}\right\rangle \left\langle \widetilde{C}_{k,l}\right\vert
\end{equation*}%
projects onto a sub-space of $P\mathcal{H\otimes H}$ spanned by the coherent
states $\left\vert \widetilde{C}_{k,l}\right\rangle $

In the other hand $(Z\overline{Z})^{-1}=(a(1-b))^{-1}$ where $a=\overline{z}%
^{P}z^{P}\otimes id_{\mathcal{H}}+id_{P\mathcal{H}}\otimes z\overline{z}$ is
a positive diagonal operators in $P\mathcal{H\otimes H}$ where the coherent
state $\left\vert \widetilde{C}_{N_{2},M_{2}}\right\rangle =\left\vert
N_{2},M_{2}\right\rangle \otimes \left\vert 0,0\right\rangle $ is projected
out and $b=a^{-1}(\overline{z}^{P}\otimes \overline{z}+z^{P}\otimes z)$. Due
to $a\left\vert \widetilde{C}_{k,l}\right\rangle =(\overline{z}%
^{P}z^{P}\otimes id_{\mathcal{H}}+id_{P\mathcal{H}}\otimes z\overline{z}%
)\left\vert \widetilde{C}_{k,l}\right\rangle =(\overline{z}^{P}\otimes 
\overline{z}+z^{P}\otimes z)\left\vert \widetilde{C}_{k,l}\right\rangle $
deduced from $Z\overline{Z}\left\vert \widetilde{C}_{k,l}\right\rangle =0$, $%
b$ acts as the unity in $P^{C}(\mathcal{H\otimes H})$ . And therefore, as
stated above, $(1-b)$ is not invertible in this domain.

In $P_{N_{1}M_{1}}(\mathcal{H\otimes H)}=(P\otimes id_{\mathcal{H}}-P^{C})(%
\mathcal{H\otimes H})$, $\left\Vert b\right\Vert \prec 1$ and $a$ is
invertible hence

\begin{equation*}
(Z\overline{Z})^{-1}=(a(1-b))^{-1}=(\sum_{n=0}^{\infty
}b^{n})a^{-1}=\sum_{k=0}^{\infty }(a^{-1}(\overline{z}^{P}\otimes \overline{z%
}+z^{P}\otimes z))^{n}a^{-1}
\end{equation*}%
which shows that $(Z\overline{Z})^{-1}:\left\vert k,l\right\rangle \otimes
\left\vert n_{1},n_{2}\right\rangle \longrightarrow P_{N_{1}M_{1}}(\mathcal{%
H\otimes H})$. Therefore

\begin{equation*}
\psi ^{1}:(P^{\top }\otimes id_{\mathcal{H}})P_{N_{1}M_{1}}(\mathcal{%
H\otimes H})\longrightarrow P_{N_{1}M_{1}}(\mathcal{H\otimes H})
\end{equation*}%
is normalized on the sub-space $P_{N_{1}M_{1}}(\mathcal{H\otimes H})$ as

\begin{equation*}
\psi ^{\dagger 1}\psi ^{1}=P_{N_{1}M_{1}}.
\end{equation*}

The same reasoning shows that the solution $\psi ^{2}$:$range(J^{\dagger
}J)=P_{\bot }\mathcal{H\otimes H}$ $\longrightarrow P\mathcal{H\otimes H}$,
where

\begin{equation*}
P_{\bot }=P_{N_{1}}+P_{M_{1}}-\left\vert N_{1},M_{1}\right\rangle
\left\langle N_{1},M_{1}\right\vert ,
\end{equation*}%
is normalized as

\begin{equation*}
\psi ^{\dagger 2}\psi ^{2}=id_{P\mathcal{H}}\otimes id_{\mathcal{H}}.
\end{equation*}

The normalization of these solutions can be recasted in

\begin{equation}
\Psi ^{\dagger }\Psi =\left( 
\begin{array}{cc}
P_{N_{1}M_{1}} & 0 \\ 
0 & id_{P\mathcal{H}}\otimes id_{\mathcal{H}}%
\end{array}%
\right) =\mathbf{P}_{N_{1}M_{1}}  \label{NORMPSI2}
\end{equation}%
where $\Psi =(\psi ^{1},\psi ^{2})$. To get normalized solutions on the full
Fock space $P\mathcal{H\otimes H}$, we proceed as follows: from the fact
that these solutions are not uniquely defined one is free to perform a $U(2)$
transformation

\begin{equation}
\widetilde{\Psi }=\Psi U
\end{equation}%
where

\begin{equation}
U=\left( 
\begin{array}{cc}
Z_{2} & Z_{1} \\ 
-\overline{Z}_{1} & \overline{Z}_{2}%
\end{array}%
\right) 
\begin{pmatrix}
(Z\overline{Z}+\Theta _{2})^{\frac{-1}{2}} & 0 \\ 
0 & (Z\overline{Z}+\Theta _{1})^{\frac{-1}{2}}%
\end{pmatrix}
\label{UNM}
\end{equation}%
with $\Theta _{1}=I_{1}I_{1}^{\dagger }-J_{1}^{\dagger }J_{1}$, $\Theta
_{2}=I_{2}I_{2}^{\dagger }-J_{2}^{\dagger }J_{2}$.

as consequence of commutation rules (\ref{COMZ}) we get

\begin{eqnarray}
U^{\dagger }U &=&%
\begin{pmatrix}
(Z\overline{Z}+\Theta _{2})^{\frac{-1}{2}} & 0 \\ 
0 & (Z\overline{Z}+\Theta _{1})^{\frac{-1}{2}}%
\end{pmatrix}%
\times 
\begin{pmatrix}
\overline{Z}_{2}Z_{2}+Z_{1}\overline{Z}_{1} & \overline{Z}_{2}Z_{1}-Z_{1}%
\overline{Z}_{2} \\ 
\overline{Z}_{1}Z_{2}-Z_{2}\overline{Z_{2}} & \overline{Z}_{1}Z_{1}+Z_{2}%
\overline{Z}_{2}%
\end{pmatrix}
\notag \\
&&\times 
\begin{pmatrix}
(Z\overline{Z}+\Theta _{2})^{\frac{-1}{2}} & 0 \\ 
0 & (Z\overline{Z}+\Theta _{1})^{\frac{-1}{2}}%
\end{pmatrix}
\notag \\
&=&%
\begin{pmatrix}
id_{P\mathcal{H}}\otimes id_{\mathcal{H}} & 0 \\ 
0 & id_{P\mathcal{H}}\otimes id_{\mathcal{H}}%
\end{pmatrix}%
=id.  \label{UCRUU}
\end{eqnarray}

Similarly we have

\begin{equation}
UU^{\dagger }=\left( 
\begin{array}{cc}
Z_{2}\frac{1}{Z\overline{Z}+\Theta _{2}}\overline{Z}_{2}+Z_{1}\frac{1}{Z%
\overline{Z}+\Theta _{1}}\overline{Z}_{1} & -Z_{2}\frac{1}{Z\overline{Z}%
+\Theta _{2}}Z_{1}+Z_{1}\frac{1}{Z\overline{Z}+\Theta _{1}}Z_{2} \\ 
-\overline{Z}_{1}\frac{1}{Z\overline{Z}+\Theta _{2}}\overline{Z}_{2}+%
\overline{Z}_{2}\frac{1}{Z\overline{Z}+\Theta _{1}}\overline{Z}_{1} & 
\overline{Z}_{1}\frac{1}{Z\overline{Z}+\Theta _{2}}Z_{1}+\overline{Z}_{2}%
\frac{1}{Z\overline{Z}+\Theta _{1}}Z_{2}%
\end{array}%
\right) .  \label{UUCRN}
\end{equation}

By using (\ref{REPN}) and (\ref{IICROI}), we show that

\begin{equation*}
\left[ z_{1}^{P},I_{2}I_{2}^{\dagger }\right] =\left[ z_{1}^{P},J_{2}^{%
\dagger }J_{2}\right] =0,\text{ \ }\left[ z_{2}^{P},I_{1}I_{1}^{\dagger }%
\right] =\left[ z_{2}^{P},J_{1}^{\dagger }J_{1}\right] =0
\end{equation*}%
from which we deduce

\begin{equation}
\left[ Z_{1},I_{2}I_{2}^{\dagger }\right] =\left[ Z_{1},J_{2}^{\dagger }J_{2}%
\right] =0,\text{ \ }\left[ Z_{2},I_{1}I_{1}^{\dagger }\right] =\left[
Z_{2},J_{1}^{\dagger }J_{1}\right] =0.  \label{ComZIIC}
\end{equation}

The commutation rules (\ref{COMZ}) lead to

\begin{equation}
\frac{1}{Z\overline{Z}}Z_{\alpha }=Z_{\alpha }\frac{1}{Z\overline{Z}+\Theta
_{\alpha }}\text{ \ , \ }\overline{Z}_{\alpha }\frac{1}{Z\overline{Z}}=\frac{%
1}{Z\overline{Z}+\Theta _{\alpha }}\overline{Z}_{\alpha }\text{.}
\label{comZinZZ}
\end{equation}

Then from (\ref{ComZIIC}) and (\ref{comZinZZ}), (\ref{UUCRN}) reads

\begin{equation}
UU^{\dagger }=\left( 
\begin{array}{cc}
\frac{1}{Z\overline{Z}}Z\overline{Z}=P_{N_{1}M_{1}} & \frac{1}{Z\overline{Z}}%
(Z_{1}Z_{2}-Z_{2}Z_{1})=0 \\ 
(\overline{Z}_{2}\overline{Z}_{1}-\overline{Z}_{1}\overline{Z}_{2})\frac{1}{Z%
\overline{Z}}=0 & \frac{1}{Z\overline{Z}+II^{\dagger }-J^{\dagger }J}%
\overline{Z}_{2}Z_{2}+\overline{Z}_{1}Z_{1}=id_{P\mathcal{H\otimes H}}%
\end{array}%
\right) =\mathbf{P}_{N_{1}M_{1}}.  \label{UUCR}
\end{equation}

The relations (\ref{UCRUU}) and (\ref{UUCR}) show that $U$ is a partial
isometry

\begin{equation}
U^{\dagger }U=id\text{ \ },\text{ \ }UU^{\dagger }=\mathbf{P}_{N_{1}M_{1}}
\label{PARTIAL}
\end{equation}%
which satisfies the relations%
\begin{equation}
\mathbf{P}_{N_{1}M_{1}}U=U\text{ and \ }U^{\dagger }\mathbf{P}%
_{N_{1}M_{1}}=U^{\dagger }  \label{UPONM}
\end{equation}%
as a consequence of $\overline{Z}_{\alpha }\left\vert \widetilde{C}%
_{k,l}\right\rangle =0$.

From\ (\ref{NORMPSI2}), (\ref{UPONM}) and (\ref{PARTIAL}) we get

\begin{equation}
\widetilde{\Psi }^{\dagger }\widetilde{\Psi }=U^{\dagger }\Psi ^{\dagger
}\Psi U=U^{\dagger }\mathbf{P}_{N_{1}M_{1}}U=U^{\dagger }U=id
\label{NORMPSIU2}
\end{equation}%
which shows that $\widetilde{\Psi }=\Psi U$ is normalized to the unity in
the full Fock space $P\mathcal{H\otimes H}$ as required by ADHM
constructions. Hence the gauge fields $\widetilde{A}^{ab}=\widetilde{\Psi }%
^{\dagger a}d\widetilde{\Psi }^{b}$ satisfying

\begin{equation}
\widetilde{A}=\widetilde{\Psi }^{\dagger }d\widetilde{\Psi }=U^{\dagger
}\Psi ^{\dagger }d(\Psi )U+U^{\dagger }\Psi ^{\dagger }\Psi dU=U^{\dagger
}\Psi ^{\dagger }d(\Psi )U+U^{\dagger }\mathbf{P}_{N_{1}M_{1}}dU=U^{\dagger
}AU+U^{\dagger }dU  \label{Atilde}
\end{equation}%
where $A=\Psi ^{\dagger }d\Psi $ and $d=P\otimes d$ is the exterior
derivative.

The field strength is given by $\widetilde{F}=d\widetilde{A}+\widetilde{A}%
\widetilde{A}$ or in terms of components by

\begin{eqnarray}
\widetilde{F}^{ab} &=&d\widetilde{A}^{ab}+\widetilde{A}^{ac}\widetilde{A}%
^{cb}=d\widetilde{\Psi }^{\dagger a}d\widetilde{\Psi }^{b}+\widetilde{\Psi }%
^{\dagger a}d\widetilde{\Psi }^{c}\widetilde{\Psi }^{\dagger c}d\widetilde{%
\Psi }^{b}  \notag \\
&=&d\widetilde{\Psi }^{\dagger a}(1-\widetilde{\Psi }^{c}\widetilde{\Psi }%
^{\dagger c})d\widetilde{\Psi }^{b}  \label{Ftild0}
\end{eqnarray}%
where we have used (\ref{NORMPSIU2}). The projector $1-\widetilde{\Psi }^{c}%
\widetilde{\Psi }^{\dagger c}$ that projects out of the zero-modes must be
equal to $\mathcal{D}_{z}^{\dagger }\frac{1}{\mathcal{D}_{z}\mathcal{D}%
_{z}^{\dagger }}\mathcal{D}_{z}$. This completeness relation, $\widetilde{%
\Psi }^{c}\widetilde{\Psi }^{\dagger c}+\mathcal{D}_{z}^{\dagger }\frac{1}{%
\mathcal{D}_{z}\mathcal{D}_{z}^{\dagger }}\mathcal{D}_{z}=1$, is necessary
to have self-dual field configurations \cite{CHU} \cite{Popov} \cite{Tatiana}
\cite{TIAN2}. To check this completeness relation, we rewrite the solutions (%
\ref{SOLZ}) under the form

\begin{equation*}
\widetilde{\Psi }^{a}=%
\begin{pmatrix}
\widetilde{\Psi }_{I}^{a} \\ 
\widetilde{\Psi }_{II}^{a}%
\end{pmatrix}%
\text{ \ where \ }\widetilde{\Psi }_{I}^{a}=%
\begin{pmatrix}
\widetilde{\Psi }_{1}^{a} \\ 
\widetilde{\Psi }_{2}^{a}%
\end{pmatrix}%
\text{ and \ }\widetilde{\Psi }_{II}^{a}=%
\begin{pmatrix}
\widetilde{\Psi }_{3}^{a} \\ 
\widetilde{\Psi }_{4}^{a}%
\end{pmatrix}%
.
\end{equation*}

Then the $4\times 4$ matrix $\widetilde{\Psi }\widetilde{\Psi }^{\dagger
}=\Psi UU^{\dagger }\Psi ^{\dagger }=\Psi \mathbf{P}_{N_{1}M_{1}}\Psi
^{\dagger }=\psi ^{1}P_{N_{1}M_{1}}\psi ^{1\dagger }+\psi ^{2}\psi
^{2\dagger }$ can be rewritten under four blocs of $2\times 2$ matrices $%
M_{A,B}=$ $\psi _{A}^{1}P_{N_{1}M_{1}}\psi _{B}^{1\dagger }+\psi
_{A}^{2}\psi _{B}^{2\dagger }$, $(A,B\in \left[ I,II\right] )$. Explicitly
we get

\begin{equation}
M_{I,I}=%
\begin{pmatrix}
-\overline{Z}_{2} \\ 
-\overline{Z}_{1}%
\end{pmatrix}%
((Z\overline{Z})^{-1}-\Delta ^{-1})%
\begin{pmatrix}
-Z_{2} & -Z_{1}%
\end{pmatrix}%
+%
\begin{pmatrix}
Z_{1} \\ 
-Z_{2}%
\end{pmatrix}%
((\overline{Z}Z)^{-1}-\Delta ^{-1})%
\begin{pmatrix}
\overline{Z}_{1} & -\overline{Z}_{2}%
\end{pmatrix}
\label{MAB}
\end{equation}%
where we have used

\begin{eqnarray*}
\chi ^{-2}P_{N_{1}M_{1}} &=&\Delta ^{-1}Z\overline{Z}(II^{\dagger
})^{-1}P_{N_{1}M_{1}}=\Delta ^{-1}Z\overline{Z}P(II^{\dagger })^{-1}= \\
&=&\Delta ^{-1}Z\overline{Z}(II^{\dagger })^{-1}=\chi ^{-2}\Longrightarrow
\chi ^{-1}P_{N_{1}M_{1}}=\chi ^{-1},
\end{eqnarray*}

deduced from $\overline{Z}_{\alpha }\left\vert \widetilde{C}%
_{k,l}\right\rangle =0$ and $\left[ II^{\dagger },P_{N_{1}M_{1}}\right] =0$,
and

\begin{eqnarray*}
(Z\overline{Z})^{-1}II^{\dagger }\chi ^{-2}II^{\dagger }(Z\overline{Z})^{-1}
&=&(Z\overline{Z})^{-1}II^{\dagger }\Delta ^{-1}=(Z\overline{Z})^{-1}-\Delta
^{-1}, \\
(\overline{Z}Z)^{-1}J^{\dagger }J\phi ^{-2}J^{\dagger }J(\overline{Z}Z)^{-1}
&=&(\overline{Z}Z)^{-1}J^{\dagger }J\Delta ^{-1}=(\overline{Z}Z)^{-1}-\Delta
^{-1}
\end{eqnarray*}%
where $\Delta =Z\overline{Z}+II^{\dagger }=\overline{Z}Z+J^{\dagger }J$.

By using (\ref{comZinZZ}), we see that the first and third terms of (\ref%
{MAB}) give%
\begin{equation*}
\begin{pmatrix}
-\overline{Z}_{2} \\ 
-\overline{Z}_{1}%
\end{pmatrix}%
(Z\overline{Z})^{-1}%
\begin{pmatrix}
-Z_{2} & -Z_{1}%
\end{pmatrix}%
+%
\begin{pmatrix}
Z_{1} \\ 
-Z_{2}%
\end{pmatrix}%
(\overline{Z}Z)^{-1}%
\begin{pmatrix}
\overline{Z}_{1} & -\overline{Z}_{2}%
\end{pmatrix}%
=1_{2\times 2}
\end{equation*}%
hence

\begin{equation}
M_{I,I}=1_{2\times 2}-%
\begin{pmatrix}
-\overline{Z}_{2} \\ 
-\overline{Z}_{1}%
\end{pmatrix}%
\Delta ^{-1}%
\begin{pmatrix}
-Z_{2} & -Z_{1}%
\end{pmatrix}%
-%
\begin{pmatrix}
Z_{1} \\ 
-Z_{2}%
\end{pmatrix}%
\Delta ^{-1}%
\begin{pmatrix}
\overline{Z}_{1} & -\overline{Z}_{2}%
\end{pmatrix}%
.  \label{MAIBI}
\end{equation}

From the relation $(Z\overline{Z})^{-1}II^{\dagger }\chi ^{-2}=\Delta ^{-1}=(%
\overline{Z}Z)^{-1}J^{\dagger }J\phi ^{-2}$ we get

\begin{equation}
M_{I,II}=%
\begin{pmatrix}
-\overline{Z}_{2} \\ 
-\overline{Z}_{1}%
\end{pmatrix}%
\Delta ^{-1}%
\begin{pmatrix}
I_{1} & I_{2}%
\end{pmatrix}%
+%
\begin{pmatrix}
Z_{1} \\ 
-Z_{2}%
\end{pmatrix}%
\Delta ^{-1}%
\begin{pmatrix}
J_{1}^{\dagger } & J_{2}^{2}%
\end{pmatrix}
\label{MAIBII}
\end{equation}%
and $M_{II,I}=M_{I,II}^{\dagger }$.

The relations $(II^{\dagger })^{-1}(Z\overline{Z})\Delta ^{-1}=(II^{\dagger
})^{-1}-\Delta ^{-1}$ and $(J^{\dagger }J)^{-1}(\overline{Z}Z)\Delta
^{-1}=(J^{\dagger }J)^{-1}-\Delta ^{-1}$ lead to

\begin{equation}
M_{II,II}=%
\begin{pmatrix}
I_{1}^{\dagger } \\ 
I_{2}^{\dagger }%
\end{pmatrix}%
(II^{\dagger })^{-1}-\Delta ^{-1})%
\begin{pmatrix}
I_{1} & I_{2}%
\end{pmatrix}%
+%
\begin{pmatrix}
J_{1}^{\dagger } \\ 
J_{2}^{\dagger }%
\end{pmatrix}%
((J^{\dagger }J)^{-1}-\Delta ^{-1})%
\begin{pmatrix}
J_{1} & J_{2}%
\end{pmatrix}%
.  \label{MAIIBII}
\end{equation}

And therefore, (\ref{MAIBI}), (\ref{MAIIBII}), (\ref{MAIBII}) and its
adjoint lead to $\widetilde{\Psi }\widetilde{\Psi }^{\dagger }=\Psi
UU^{\dagger }\Psi ^{\dagger }=\Psi P_{N_{1}M_{1}}\Psi ^{\dagger }=1_{\Psi }-%
\mathcal{D}_{z}^{\dagger }\frac{1}{\mathcal{D}_{z}\mathcal{D}_{z}^{\dagger }}%
\mathcal{D}_{z}$ where

\begin{equation}
1_{\Psi }=\left( 
\begin{array}{cc}
1_{2\times 2} & 0_{2\times 2} \\ 
0_{2\times 2} & 
\begin{pmatrix}
I_{1}^{\dagger } \\ 
I_{2}^{\dagger }%
\end{pmatrix}%
(II^{+})^{-1}%
\begin{pmatrix}
I_{1} & I_{2}%
\end{pmatrix}%
+%
\begin{pmatrix}
J_{1} \\ 
J_{2}%
\end{pmatrix}%
(J^{+}J)^{-1}%
\begin{pmatrix}
J_{1}^{\dagger } & J_{2}^{\dagger }%
\end{pmatrix}%
\end{array}%
\right)  \label{FERMETURE}
\end{equation}%
which is the unit for the zero mode $\widetilde{\Psi }^{a}$. Due to $%
dI_{\alpha }=I_{\alpha }d$ and $dJ_{\alpha }=J_{\alpha }d$ (\ref{FERMETURE})
is also the unit for $d\widetilde{\Psi }^{a}$. Hence the completeness
relation from which the field strength components (\ref{Ftild0}) reads

\begin{eqnarray}
\widetilde{F}^{ab} &=&d\widetilde{\Psi }^{\dagger a}\mathcal{D}_{z}^{\dagger
}\frac{1}{\mathcal{D}_{z}\mathcal{D}_{z}^{\dagger }}\mathcal{D}_{z}d%
\widetilde{\Psi }^{b}=\widetilde{\Psi }^{\dagger a}(d\mathcal{D}%
_{z}^{\dagger })\frac{1}{\mathcal{D}_{z}\mathcal{D}_{z}^{\dagger }}(d%
\mathcal{D}_{z})\widetilde{\Psi }^{b}  \label{Ftild} \\
&=&U^{\dagger ac}\psi ^{\dagger c}(d\mathcal{D}_{z}^{\dagger })\frac{1}{%
\mathcal{D}_{z}\mathcal{D}_{z}^{\dagger }}(d\mathcal{D}_{z})\psi
^{d}U^{db}=U^{\dagger ac}F^{cd}U^{db}.  \notag
\end{eqnarray}%
where

\begin{eqnarray}
F^{ab} &=&\psi ^{\dagger a}(d\mathcal{D}_{z}^{\dagger })\frac{1}{\mathcal{D}%
_{z}\mathcal{D}_{z}^{\dagger }}(d\mathcal{D}_{z})\psi ^{b}  \notag \\
&=&\psi ^{\dagger }\frac{1}{\Delta }%
\begin{pmatrix}
dz_{1}d\overline{z}_{1}-dz_{2}d\overline{z}_{2} & -2dz_{1}d\overline{z}_{2}
& 0 & 0 \\ 
-2dz_{2}d\overline{z}_{1} & dz_{2}d\overline{z}_{2}-dz_{1}d\overline{z}_{1}
& 0 & 0 \\ 
0 & 0 & 0 & 0 \\ 
0 & 0 & 0 & 0%
\end{pmatrix}%
\psi ^{b}.  \label{FAB}
\end{eqnarray}

The relation (\ref{FAB}) exhibits the anti-self duality conditions 
\begin{equation*}
F_{z_{1}\overline{z}_{1}}^{ab}=-F_{z_{2}\overline{z}_{2}}^{ab},\text{ \ \ \ }%
F_{z_{1}z_{2}}^{ab}=0\Longrightarrow \widetilde{F}_{z_{1}\overline{z}%
_{1}}^{ab}=-\widetilde{F}_{z_{2}\overline{z}_{2}}^{ab},\text{ \ \ }%
\widetilde{F}_{z_{1}z_{2}}^{ab}=0
\end{equation*}%
and the $U(2)$ instanton number is given, in this formalism, by%
\begin{equation}
k=-\frac{(4\pi \theta )^{2}}{8\pi ^{2}}Tr(\widetilde{F})^{2}
\label{kinstftild}
\end{equation}%
where now $Tr=tr_{P\mathcal{H}}\otimes tr_{\mathcal{H}}tr_{U(2)}$.

Let us now consider the particular case by setting $N_{1}=M_{1}=0$ into the
projector (\ref{PROJ}). From (\ref{I1J1ADHM}) we deduce $J_{1}=J_{2}=0$
which reduces the solution (\ref{SOLZ}) to $\psi =(\psi ^{1},0)$ where $\psi
^{1}$ is normalized as

\begin{equation*}
\psi ^{1\dagger }\psi ^{1}=P_{00}
\end{equation*}%
with

\begin{equation*}
P_{00}=P\otimes id_{\mathcal{H}}-\sum_{k=0}^{N_{2}}\sum_{l=0}^{M_{2}}\left%
\vert \widetilde{C}_{kl}\right\rangle \left\langle \widetilde{C}%
_{kl}\right\vert .
\end{equation*}

Since $I_{\alpha }P=0$ The solution $\psi _{2}=0$ may be replaced by

\begin{equation}
\psi ^{2}=\left( 
\begin{array}{c}
0 \\ 
0 \\ 
\frac{1}{\sqrt{2}}P\otimes id_{\mathcal{H}} \\ 
\frac{1}{\sqrt{2}}P\otimes id_{\mathcal{H}}%
\end{array}%
\right)  \label{PSITRIVIAL}
\end{equation}%
which is a trivial solution to (\ref{DIRACZ}) with $J_{1}^{\dagger
}=J_{2}^{\dagger }=0$, orthogonal to $\psi ^{1}$ and normalized as $\psi
^{2\dagger }\psi ^{2}=id_{P\mathcal{H\otimes H}}$. This solution can be
recasted in $\Psi =(\psi ^{1},\psi ^{2})$ normalized as

\begin{equation}
\Psi ^{\dagger }\Psi =\left( 
\begin{array}{cc}
P_{00} & 0 \\ 
0 & id_{P\mathcal{H\otimes H}}%
\end{array}%
\right) =\mathbf{P}_{00}.  \label{NORMPSI1}
\end{equation}

From that, it is obvious to proceed as for the $U(2)$ solutions above by
transforming the doublet $\Psi $ by the partial isometry (\ref{UNM}) with $%
\Theta _{1}=I_{1}I_{1}^{\dagger }$ and $\Theta _{2}=I_{2}I_{2}^{\dagger }$
to get

\begin{equation*}
\widetilde{\Psi }^{\dagger }\Psi =U^{\dagger }\Psi ^{\dagger }\Psi
U=U^{\dagger }\mathbf{P}_{00}U=U^{\dagger }U=id.
\end{equation*}

The gauge field $\widetilde{A}=\widetilde{\Psi }^{\dagger }d\widetilde{\Psi }
$ transforms like (\ref{Atilde}) with $A^{12}=\psi ^{1\dagger }d\psi ^{2}=0$%
, $A^{21}=\psi ^{2\dagger }d\psi ^{1}=0$ and $A^{22}=\psi ^{2\dagger }d\psi
^{2}=0$. The same technic can be used to check that $\widetilde{\Psi }%
\widetilde{\Psi }^{\dagger }=1_{\Psi }-\mathcal{D}_{z}^{\dagger }\frac{1}{%
\mathcal{D}_{z}\mathcal{D}_{z}^{\dagger }}\mathcal{D}_{z}$ with

\begin{equation*}
1_{\Psi }=\left( 
\begin{array}{cc}
1_{2\times 2} & 0_{2\times 2} \\ 
0_{2\times 2} & 
\begin{pmatrix}
I_{1}^{\dagger } \\ 
I_{2}^{\dagger }%
\end{pmatrix}%
(II^{+})^{-1}%
\begin{pmatrix}
I_{1} & I_{2}%
\end{pmatrix}%
+\frac{1}{2}\left( 
\begin{array}{cc}
P\otimes id_{\mathcal{H}} & P\otimes id_{\mathcal{H}} \\ 
P\otimes id_{\mathcal{H}} & P\otimes id_{\mathcal{H}}%
\end{array}%
\right)%
\end{array}%
\right)
\end{equation*}%
which is the unit of $\ \widetilde{\Psi }^{a}$ and $d\widetilde{\Psi }^{a}$.
Then the field strength components can be computed from (\ref{FAB}) to get $%
F^{12}=0$, $F^{21}=0$, $F^{22}=0$ but all the components of $\widetilde{F}%
=U^{\dagger }FU$ are nonvanishing and are given in terms of $F^{11}$, $%
Z_{\alpha }$ and $\overline{Z}_{\alpha }$. This shows that, in this
formalism, the solution $\psi ^{1}$ which is generally considered as
solutions for $U(1)$ instantons is in fact a doublet component of $U(2)$
instantons.

Let us now investigate a little more the ADHM constraints. The above results
show that depending the form of the projector (\ref{PROJ}), we obtain
different solutions of the ADHM constraints. These solutions depend on the
form of $z_{\alpha }^{P}$ which determine the form of the operators $%
I_{\alpha }$ and $J_{\alpha }$ in (\ref{ADHMz}). In fact from (\ref{zppz}), (%
\ref{I1J1ADHM}), $PI_{\alpha }^{\dagger }=0=PJ_{\alpha }$, $PI_{\alpha
}=I_{\alpha }$ and $PJ_{\alpha }^{\dagger }=J_{\alpha }^{\dagger }$ we deduce

\begin{equation}
Pz_{\alpha }P=Pz_{\alpha }-J_{\alpha }^{\dagger }=z_{\alpha }^{P}\text{ \ \
, \ \ }z_{\alpha }P=Pz_{\alpha }P+I_{\alpha }^{\dagger }=z_{\alpha
}^{P}+I_{\alpha }^{\dagger }  \label{ppzzp}
\end{equation}%
which show that for $z_{2}^{P}=0$, we get $z_{2}P=I_{2}^{\dagger }$ and $%
Pz_{2}=J_{2}^{\dagger }$, and for $z_{1}^{P}=0$ we get $z_{1}P=I_{1}^{%
\dagger }$ and $Pz_{\alpha }=J_{\alpha }^{\dagger }$. These cases are
respectivelly obtained from projectors of the form

\begin{equation*}
P_{2s}=\sum_{n_{1}=0}^{N_{2}}\left\vert n_{1}+k,l\right\rangle \left\langle
n_{1}+k,l\right\vert \text{ and }P_{1s}=\sum_{n_{2}=0}^{M_{2}}\left\vert
k,n_{2}+l\right\rangle \left\langle k,n_{2}+l\right\vert \text{ \ }\forall
k,l\geq 0.
\end{equation*}

$k=l=0$ corresponds to $N_{1}=M_{1}=0\Longrightarrow J_{1}=J_{2}=0$.

For $z_{1}^{P}=z_{2}^{P}=0$ we get $z_{\alpha }P=I_{\alpha }^{\dagger }$ and 
$Pz_{\alpha }=J_{\alpha }^{\dagger }$. This case is obtained from projectors
of the form

\begin{equation}
P=\sum_{n_{1}=N_{1}}^{N_{2}}\left\vert n_{1}+k,n_{1}+l\right\rangle
\left\langle n_{1}+k,n_{1}+l\right\vert \text{ \ }\forall k,l\geq 0\text{ \ }%
or\text{ \ }P_{J}=\sum_{m=-J}^{J}\left\vert J,m\right\rangle \left\langle
J,m\right\vert  \label{TRANLAPROJ}
\end{equation}%
where $J=\frac{1}{2}(n_{1}+n_{2})=0,\frac{1}{2},1,...\infty \,$and $m=\frac{1%
}{2}(n_{1}-n_{2})$\ runs by integer steps over the range $-$ $J\leq m\leq J$.

Finally if we take the projector

\begin{equation*}
P^{bd}=P^{\top }+P_{\bot }-P_{N_{1}M_{2}}-P_{N_{2}M_{1}}\text{ \ with }%
P_{N_{\alpha }M_{\beta }}=\left\vert N_{\alpha },M_{\beta }\right\rangle
\left\langle N_{\alpha },M_{\beta }\right\vert ,\text{ }(\alpha ,\beta \in %
\left[ 1,2\right] )
\end{equation*}%
which projects onto the boundary of $P\mathcal{H}$ where $P$ is given by (%
\ref{PROJ}), we get

\begin{equation}
\lbrack \overline{z}_{\alpha }^{bd},z_{\beta }^{bd}]=2\theta \delta _{\alpha
\beta }-(I_{\alpha }^{bd}I_{\beta }^{bd\dagger }-J_{\alpha }^{bd\dagger
}J_{\beta }^{bd})  \label{comzpbd}
\end{equation}%
where $z_{\alpha }^{bd}=P^{bd}z_{\alpha }P^{bd}$ and

\begin{eqnarray*}
I_{1}^{bd} &=&(P_{N_{2}}+P_{N_{1}}-P_{N_{1}M_{1}}-P_{N_{1}M_{2}})\overline{z}%
_{1},\text{ \ }I_{2}^{bd}=(P_{M_{2}}+P_{M_{1}}-P_{N_{1}M_{1}}-P_{N_{2}M_{1}})%
\overline{z}_{2}, \\
J_{1}^{bd} &=&\overline{z}%
_{1}(P_{N1}+P_{N_{2}}-P_{N_{2}M_{1}}-P_{N_{2}M_{2}}),\text{ \ }J_{2}^{bd}=%
\overline{z}_{2}(P_{M_{1}}+P_{M_{2}}-P_{N_{1}M_{2}}-P_{N_{2}M_{2}}).
\end{eqnarray*}

The relations (\ref{comzpbd}) give solutions of the real ADHM constraint (%
\ref{ADHMz}) with $I^{bd}=\left( I_{1}^{bd},I_{2}^{bd}\right) $ and $%
J^{bd\dagger }=\left( J_{2}^{bd\dagger },-J_{1}^{bd\dagger }\right) $ and
the commutation relation

\begin{equation*}
\lbrack \overline{z}_{1}^{bd},\overline{z}%
_{2}^{bd}]=I_{2}^{bd}J_{1}^{bd}-I_{1}^{bd}J_{2}^{bd}=\overline{z}%
_{1}P_{N_{2}M_{1}}\overline{z}_{2}-\overline{z}_{2}P_{N_{1}M_{2}}\overline{z}%
_{1}
\end{equation*}%
corresponds to the complex ADHM constraint $\left[ B_{1},B_{2}\right] +IJ=0.$

The same definitions of $Z_{\alpha }$, $I_{\alpha }$ and $J_{\alpha }$ given
below (\ref{DIRACZ}) in term of the projector $P^{bd}$ lead to

\begin{equation}
\lbrack Z_{\alpha }^{bd},\overline{Z}_{\beta }^{bd}]=-(I_{\alpha
}^{bd}I_{\beta }^{bd\dagger }-J_{\alpha }^{bd\dagger }J_{\beta }^{bd}),\text{
}[Z_{1}^{bd},Z_{2}^{bd}]=I_{2}^{bd}J_{1}^{bd}-I_{1}^{bd}J_{2}^{bd}.
\label{COMZBD}
\end{equation}

The Dirac operator which is compatible with these commutation relations reads

\begin{equation*}
\mathcal{D}_{z}=\left( 
\begin{array}{c}
\begin{array}{ccc}
Z_{2} & Z_{1} & I%
\end{array}
\\ 
\begin{array}{ccc}
-\overline{Z}_{1} & \overline{Z}_{2} & J^{\dagger }%
\end{array}%
\end{array}%
\right) =\left( 
\begin{array}{c}
\begin{array}{cccc}
Z_{2} & Z_{1} & I_{1} & I_{2}%
\end{array}
\\ 
\begin{array}{cccc}
-\overline{Z}_{1} & \overline{Z}_{2} & J_{2}^{\dagger } & -J_{1}^{\dagger }%
\end{array}%
\end{array}%
\right)
\end{equation*}%
leading to an invertible $\mathcal{D}_{z}\mathcal{D}_{z}^{\dagger }$
necessary to ADHM construction. In fact by using (\ref{COMZBD}) we get

\begin{equation*}
(\mathcal{D}_{z}\mathcal{D}_{z}^{\dagger })^{-1}=(Z\overline{Z}+II^{\dagger
})^{-1}1_{2\times 2}=(\overline{Z}Z+J^{\dagger }J)^{-1}1_{2\times 2}.
\end{equation*}

Because of the noncommutativity of $Z_{1}$ with $Z_{2}$ and $\overline{Z}%
_{2} $ it is more difficult to find solutions of $\mathcal{D}_{z}\psi =0$.
All the above examples of ADHM\ constraint solutions reinforce the
interpretation of $B_{1}^{\dagger }=z_{1}^{p},B_{2}^{\dagger }=z_{2}^{p}\in
End(V)=End(P\mathcal{H)}$ as instantons positions in the noncommutative
space $\mathbb{R}_{\theta }^{4}$. The noncommutative analogue of the
localized feature of positions is expressed by projectors which project onto
finite dimensional sub-space $P\mathcal{H}$ \ of the Fock space
representation of the algebra $\mathcal{A}_{\theta }(\mathbb{R}^{4})$.

Before to compute explicitly the $U(2)$ instanton number (\ref{kinstftild}),
let us note that many of the above results present similarities with works
treating noncommutative ADHM constructions of instantons. Especially the
analogy of this formalism with the algebraic-geometric interpretation of the
space $V$ and the triple $(B_{1},B_{2},I)$ considered in \cite{nekrasov1}, 
\cite{nekrasov2} and \cite{nekrasov3}. This analogy is given by the
correspondence between the Fock space $\mathcal{H}$ and the space $%
\mathbb{C}
\left[ z_{1},z_{2}\right] $ of all polynomials in classical variable $z_{1}$
and $z_{2}$ as

\begin{eqnarray*}
\left\vert n_{1},n_{2}\right\rangle &=&(z_{1}/\sqrt{2\theta }%
)^{n_{1}}/(n_{1}!)(z_{2}/\sqrt{2\theta })^{n2}/(n_{2}!)\left\vert
0,0\right\rangle \text{ }\Leftrightarrow \text{ }(z_{1})^{n_{1}}(z_{2})^{n2}
\\
P\mathcal{H} &\mathcal{=}&%
\mathbb{C}
\left[ z_{1}^{P},z_{2}^{P}\right] \left\vert 0,0\right\rangle
((z_{1}^{P})^{N_{2}+1}=0,(z_{2}^{P})^{M_{2}+1}=0)\Leftrightarrow 
\mathbb{C}
\left[ z_{1},z_{2}\right] /I_{p}=V
\end{eqnarray*}%
where $I_{p}$ is the ideal parameterizes the torsion free sheaf on $%
\mathbb{C}
^{2}$. In this case $I_{p}$ is the ideal given by the space of functions of
the form

\begin{equation*}
I_{p}=(z_{1})^{N_{2}+1}g(z_{1},z_{2})+(z_{2})^{M_{2}+1}h(z_{1},z_{2})\simeq P%
\mathcal{H}
\end{equation*}%
where $P$ is the projector (\ref{PROJ}) with $N_{1}=M_{1}=0$. For instance,
the case $N_{1}=M_{1}=M_{2}=0$,

\begin{equation*}
P=\sum_{n_{1}=0}^{N_{2}}\left\vert n_{1},0\right\rangle \left\langle
n_{1},0\right\vert \text{ ,}
\end{equation*}%
gives $I_{2}=\sqrt{2\theta }\sum_{n_{1}=0}^{N_{2}}\left\vert
n_{1},0\right\rangle \left\langle n_{1},1\right\vert $, $I_{1}=\sqrt{2\theta
(N_{2}+1)}\left\vert N_{2},0\right\rangle \left\langle N_{2}+1,0\right\vert $%
, $B_{2}=\overline{z}_{2}^{P}=0$, and $B_{1}=\overline{z}_{1}^{P}$. This
case corresponds exactly to the one's given in section (4.1) of \cite%
{nekrasov1} for $N=N_{2}-1$ where $%
I_{p}=(z_{1})^{N_{2}+1}g(z_{1},z_{2})+(z_{2})h(z_{1},z_{2})$, $B_{2}=%
\overline{z}_{2}^{P}=0$ but with $B_{1}=\sqrt{2}\overline{z}_{1}^{P}$ and
only the image of the operator $I=\sqrt{2}(I_{1}=\sqrt{2\theta N}\left\vert
N,0\right\rangle \left\langle N+1,0\right\vert )=\sqrt{4\theta N}\left\vert
N,0\right\rangle $ is taken into account. The factor $\sqrt{2}$ is added by
hand to fit to the ADHM constraints. This kind of solution is also
considered in \cite{Sako1} and, with an adequate shift of the deformation
parameter, in \cite{TIAN2}. This consists in fact to take as ADHM equations
only the sector $z_{1}^{P}-\overline{z}_{1}^{P}$ in (\ref{comzp}). This
restriction does not take into account the operator $I_{2}$ which exists
even if $z_{2}^{P}=0$ and consider, in general, only the absolute value of
the operator $I_{1}$ under the form $I=\sqrt{2}\sqrt{I_{1}I_{1}^{\dagger }}$
in the Dirac operator. Because of $IP=\sqrt{2}\sqrt{I_{1}I_{1}^{\dagger }}%
P=I $, this latter restriction does not allow to get a trivial solution of
the form (\ref{PSITRIVIAL}) and then to construct doublets of $U(2)$
instantons.

In the following, we compute explicitly the case where the projector is of
the form

\begin{equation}
P_{J}=\sum_{m=-J}^{J}\left\vert J,m\right\rangle \left\langle J,m\right\vert
.  \label{PROJM}
\end{equation}

The coordinates $z_{\alpha }$ and $\overline{z}_{\alpha }$ act on the states 
$\left\vert J,m\right\rangle $ as

\begin{eqnarray}
z_{1}\left\vert J,m\right\rangle &=&\sqrt{2\theta (J+m+1)}\left\vert J+\frac{%
1}{2},m+\frac{1}{2}\right\rangle ,  \notag \\
z_{2}\left\vert J,m\right\rangle &=&\sqrt{2\theta (J-m+1)}\left\vert J+\frac{%
1}{2},m-\frac{1}{2}\right\rangle ,  \notag \\
\overline{z}_{1}\left\vert J,m\right\rangle &=&\sqrt{2\theta (J-m)}%
\left\vert J-\frac{1}{2},m-\frac{1}{2}\right\rangle ,  \notag \\
\overline{z}_{2}\left\vert J,m\right\rangle &=&\sqrt{2\theta (J-m)}%
\left\vert J-\frac{1}{2},m+\frac{1}{2}\right\rangle ,  \notag \\
z\overline{z}\left\vert J,m\right\rangle &=&4\theta J\left\vert
J,m\right\rangle  \label{REPJM}
\end{eqnarray}%
which imply that $z_{\alpha }^{P_{J}}=0$, $\overline{z}_{\alpha }^{P_{J}}=0$
and

\begin{eqnarray}
I_{1} &=&P_{J}\overline{z}_{1}\ ,\text{ }I_{2}=P_{J}\overline{z}_{2}\text{ },%
\text{ }J_{1}=\overline{z}_{1}P_{J}\text{ },\text{ }J_{2}=\overline{z}%
_{2}P_{J},  \notag \\
I_{1}I_{1}^{\dagger } &=&2\theta \sum_{m=-J}^{J}(J+m+1)P_{Jm}\text{ },\text{ 
}I_{2}I_{2}^{\dagger }=2\theta \sum_{m=-J}^{J}(J-m+1)P_{Jm},  \notag \\
J_{1}^{\dagger }J_{1} &=&2\theta \sum_{m=-J}^{J}(J+m)P_{Jm}\text{ },\text{ }%
J_{2}^{\dagger }J_{2}=2\theta \sum_{m=-J}^{J}(J-m)P_{Jm}  \label{zPJ}
\end{eqnarray}%
where $P_{Jm}=\left\vert J,m\right\rangle \left\langle J,m\right\vert $. $%
I_{1}$,$I_{2}:P_{J+\frac{1}{2}}\mathcal{H\rightarrow }P_{J}\mathcal{H}$ and $%
J_{1}$, $J_{2}:P_{J}\mathcal{H\rightarrow }P_{J-\frac{1}{2}}\mathcal{H}$.

$z_{\alpha }^{P_{J}}=0$ reduce the ADHM conditions (\ref{ADHMz}) to $%
II^{\dagger }-J^{\dagger }J=4\theta P_{J}$ and $IJ=0$ which are easily
verified from (\ref{zPJ}). The Dirac operator (\ref{DIRACZ}) reduces to

\begin{equation*}
\mathcal{D}_{z}=\left( 
\begin{array}{c}
\begin{array}{cccc}
-z_{2} & -z_{1} & I_{1} & I_{2}%
\end{array}
\\ 
\begin{array}{cccc}
\overline{z}_{1} & -\overline{z}_{2} & J_{1}^{\dagger } & J_{2}^{\dagger }%
\end{array}%
\end{array}%
\right) ,
\end{equation*}%
where $z_{\alpha }=id_{P_{J}\mathcal{H}}\otimes z_{\alpha }$, $I_{\alpha
}=I_{\alpha }\otimes id_{\mathcal{H}}$ and $J_{\alpha }=J_{\alpha }\otimes
id_{\mathcal{H}}$. The solutions $\mathcal{D}_{z}\psi ^{a}=0$ read

\begin{equation}
\psi ^{1}=\left( 
\begin{array}{c}
II^{\dagger }\overline{z}_{2}\frac{1}{z\overline{z}} \\ 
II^{\dagger }\overline{z}_{1}\frac{1}{z\overline{z}} \\ 
I_{1}^{\dagger } \\ 
I_{2}^{\dagger }%
\end{array}%
\right) \chi ^{-1}\text{ and \ }\psi ^{2}=\left( 
\begin{array}{c}
-J^{\dagger }Jz_{1}\frac{1}{\overline{z}z} \\ 
J^{\dagger }Jz_{2}\frac{1}{\overline{z}z} \\ 
J_{1} \\ 
J_{2}%
\end{array}%
\right) \phi ^{-1}  \label{SOLUTJ}
\end{equation}%
where $\chi ^{2}=II^{\dagger }(z\overline{z})^{-1}(z\overline{z}+II^{\dagger
})$, $\phi ^{2}=J^{\dagger }J(\overline{z}z)^{-1}(\overline{z}z+J^{\dagger
}J)$ and $z\overline{z}=z_{1}\overline{z}_{1}+z_{2}\overline{z}_{2}$. The
components of $\psi ^{a}$ act on the Fock space $P_{J}\mathcal{H\otimes H}$
and are normalized as

\begin{equation}
\Psi ^{\dagger }\Psi =\left( 
\begin{array}{cc}
p_{00} & 0 \\ 
0 & id_{P_{J}\mathcal{H\otimes H}}%
\end{array}%
\right) =P_{00}  \label{NORMPSI}
\end{equation}%
where $\Psi =(\psi ^{1},\psi ^{2})$ and $p_{00}=P_{J}\otimes (id_{\mathcal{H}%
}-\left\vert 0,0\right\rangle \left\langle 0,0\right\vert )$. Since in this
case $\Theta _{1}=I_{1}I_{1}^{\dagger }-J_{1}^{\dagger }J_{1}=2\theta
id_{P_{J}\mathcal{H\otimes H}}=\Theta _{2}=I_{2}I_{2}^{\dagger
}-J_{2}^{\dagger }J_{2}$, the partial isometry (\ref{UNM}) reduces to

\begin{equation*}
U=\left( 
\begin{array}{cc}
-z_{2} & -z_{1} \\ 
\overline{z}_{1} & -\overline{z}_{2}%
\end{array}%
\right) \frac{1}{\sqrt{z\overline{z}+2\theta }}
\end{equation*}%
which satisfies the relations

\begin{equation}
U^{\dagger }U=\left( 
\begin{array}{cc}
id_{P_{J}\mathcal{H\otimes H}} & 0 \\ 
0 & id_{P_{J}\mathcal{H\otimes H}}%
\end{array}%
\right) \text{ \ , \ }UU^{\dagger }=P_{00}\text{ }  \label{UUCROIX}
\end{equation}%
and

\begin{equation}
P_{00}U=U\text{ \ , \ }U^{\dagger }P_{00}=U^{\dagger }.  \label{UPO}
\end{equation}

Note that in the classical limit ($\theta =0$)

\begin{equation}
U=\left( 
\begin{array}{cc}
-z_{2} & -z_{1} \\ 
\overline{z}_{1} & -\overline{z}_{2}%
\end{array}%
\right) \frac{1}{\sqrt{z\overline{z}}}=-id_{k}\otimes (x_{4}+ix_{i}\sigma
_{i})\frac{1}{r}\in id_{k}\otimes SU(2)  \label{USU2}
\end{equation}%
may be viewed as an element of the classical group $SU(2)$. Here $\sigma
_{i} $ are the Pauli matrices and $r=\sqrt{%
x_{1}^{2}+x_{2}^{2}+x_{3}^{2}+x_{4}^{2}}$.

From (\ref{UPO}) and (\ref{UUCROIX}) we obtain the normalization of $%
\widetilde{\Psi }=\Psi U$ as

\begin{equation}
\widetilde{\Psi }^{\dagger a}\widetilde{\Psi }^{b}=\delta ^{ab}id_{P_{J}%
\mathcal{H\otimes H}}  \label{NORMPSIT}
\end{equation}%
and the gauge field $\widetilde{A}$ satisfies the relation (\ref{Atilde}).
The components of the gauge field $A$ can be explicitly calculated\ by%
\begin{equation}
A^{ab}=\psi ^{\dagger a}\frac{1}{2\theta }\left[ \overline{z}^{i},\psi ^{b}%
\right] dz^{i}-\psi ^{\dagger a}\frac{1}{2\theta }\left[ z^{i},\psi ^{b}%
\right] d\overline{z}^{i}.
\end{equation}

In fact from $\left[ z_{\alpha },I_{\beta }I_{\beta }^{\dagger }\right] =0$
and $\overline{z}_{\alpha }f(z\overline{z})=f(z\overline{z}+2\theta id_{P_{J}%
\mathcal{H\otimes H}})\overline{z}_{\alpha }=f(z\overline{z}+2\theta )%
\overline{z}_{\alpha }$, a straightforward calculus gives

\begin{eqnarray}
A^{11} &=&\frac{1}{2\theta }((\frac{z\overline{z}(z\overline{z}+II^{\dagger
}+2\theta )}{(z\overline{z}+2\theta )(z\overline{z}+II^{\dagger })})^{\frac{1%
}{2}}-p_{00})\overline{z}_{\alpha }dz^{\alpha }-h.c\text{ , }  \notag \\
\text{\ }A^{22} &=&\frac{1}{2\theta }\overline{z}_{\alpha }((\frac{\overline{%
z}z(\overline{z}z+J^{\dagger }J-2\theta )}{(\overline{z}z-2\theta )(%
\overline{z}z+J^{\dagger }J)})^{\frac{1}{2}}-1)dz^{\alpha }-h.c.,  \notag \\
A^{12} &=&(\frac{II^{\dagger }J^{\dagger }J}{(z\overline{z})(z\overline{z}%
+II^{\dagger })(z\overline{z}+2\theta )(z\overline{z}+II^{\dagger }-2\theta )%
})^{\frac{1}{2}}(z_{1}dz_{2}-z_{2}dz_{1})\text{ , }  \notag \\
A^{21} &=&-(A^{12})^{\dagger }  \label{Aexplicite}
\end{eqnarray}%
and%
\begin{eqnarray}
U^{\dagger }dU &=&\frac{1}{2\theta }%
\begin{pmatrix}
((\frac{z\overline{z}+2\theta }{z\overline{z}+4\theta })^{\frac{1}{2}}-1)%
\overline{z}_{1} & 2\theta ((z\overline{z}+2\theta )(z\overline{z}+4\theta
))^{-\frac{1}{2}}\overline{z}_{2} \\ 
0 & ((\frac{z\overline{z}+4\theta }{z\overline{z}+2\theta })-1)\overline{z}%
_{1}%
\end{pmatrix}%
dz_{1}  \notag \\
&&+\frac{1}{2\theta }%
\begin{pmatrix}
((\frac{z\overline{z}+4\theta }{z\overline{z}+2\theta })^{\frac{1}{2}}-1)%
\overline{z}_{2} & 0 \\ 
2\theta ((z\overline{z}+2\theta )(z\overline{z}+4\theta ))^{-\frac{1}{2}}%
\overline{z}_{1} & ((\frac{z\overline{z}+2\theta }{z\overline{z}+4\theta }%
)-1)\overline{z}_{2}%
\end{pmatrix}%
dz_{2}  \notag \\
&&-\frac{1}{2\theta }%
\begin{pmatrix}
z_{1}((\frac{z\overline{z}+2\theta }{z\overline{z}+4\theta })^{\frac{1}{2}%
}-1) & 0 \\ 
2\theta z_{2}((z\overline{z}+2\theta )(z\overline{z}+4\theta ))^{-\frac{1}{2}%
} & z_{1}((\frac{z\overline{z}+4\theta }{z\overline{z}+2\theta })-1)%
\end{pmatrix}%
d\overline{z}_{1}  \notag \\
&&-\frac{1}{2\theta }%
\begin{pmatrix}
z_{2}((\frac{z\overline{z}+4\theta }{z\overline{z}+2\theta })^{\frac{1}{2}%
}-1) & 2\theta z_{1}((z\overline{z}+2\theta )(z\overline{z}+4\theta ))^{-%
\frac{1}{2}} \\ 
0 & z_{2}((\frac{z\overline{z}+2\theta }{z\overline{z}+4\theta })-1)%
\end{pmatrix}%
d\overline{z}_{2}.  \label{UCDU}
\end{eqnarray}

We get the explicit form of the field strength components $F^{ab}$ from (\ref%
{FAB}) as:

\begin{eqnarray}
F^{11} &=&-D^{11}(z\overline{z})(2\overline{z}_{1}z_{2}dz_{1}d\overline{z}%
_{2}+2z_{1}\overline{z}_{2}dz_{2}d\overline{z}_{1}+  \notag \\
&&(z_{1}\overline{z}_{1}-z_{2}\overline{z}_{2})dz_{1}d\overline{z}_{1}-(z_{1}%
\overline{z}_{1}-z_{2}\overline{z}_{2})dz_{2}d\overline{z}_{2}),  \notag \\
F^{22} &=&D^{22}(z\overline{z})(2\overline{z}_{1}z_{2}dz_{1}d\overline{z}%
_{2}+2z_{1}\overline{z}_{2}dz_{2}d\overline{z}_{1}+  \notag \\
&&(z_{1}\overline{z}_{1}-z_{2}\overline{z}_{2})dz_{1}d\overline{z}_{1}-(z_{1}%
\overline{z}_{1}-z_{2}\overline{z}_{2})dz_{2}d\overline{z}_{2}),  \notag \\
F^{12} &=&2D^{12}(z\overline{z})(-z_{2}z_{2}dz_{1}d\overline{z}%
_{2}+z_{1}z_{1}dz_{2}d\overline{z}_{1}-z_{1}z_{2}dz_{1}d\overline{z}%
_{1}+z_{1}z_{2}dz_{2}d\overline{z}_{2}),  \notag \\
F^{21} &=&(F^{12})^{\dagger }  \label{Fexplicite}
\end{eqnarray}%
where $D^{11}(z\overline{z})=\frac{II^{\dagger }}{\widehat{z}\overline{z}(z%
\overline{z}+II^{\dagger }-2\theta )(z\overline{z}+II^{\dagger })}$ , $%
D^{22}(z\overline{z})=\frac{J^{\dagger }J}{\overline{z}z(\overline{z}%
z+J^{\dagger }J+2\theta )(\overline{z}z+J^{\dagger }J)}$, $D^{12}(z\overline{%
z})=(\frac{II^{\dagger }J^{\dagger }J}{(z\overline{z}+II^{\dagger })(z%
\overline{z}+J^{\dagger }J)})^{\frac{1}{2}}\frac{1}{z\overline{z}(z\overline{%
z}+II^{\dagger }-2\theta )}$ and $(dz_{\alpha }d\overline{z}_{\beta
})^{\dagger }=dz_{\beta }d\overline{z}_{\alpha }$.

Notice that in the particular case where the projector $P_{0}$ is of the
form $P_{0}=\left\vert 0,0\right\rangle \left\langle 0,0\right\vert $, we
see from (\ref{zPJ}) that $II^{\dagger }=4\theta id_{P_{0}\mathcal{H\otimes H%
}}=4\theta $ and $\ J_{\alpha }=0$ which imply, from (\ref{Aexplicite}) and (%
\ref{Fexplicite}), that $A^{12}=A^{21}=A^{22}=0$ and $F^{12}=F^{21}=F^{22}=0 
$. The component $F^{11}$ takes the same form as for the $U(1)$
one-instanton calculated in \cite{furuuchi1}. Let us recall that $F^{11}$ is
defined in the Fock space $P_{0}\mathcal{H\otimes H\approx H}$ where the
state $\left\vert 0,0\right\rangle $ is projected out while $\widetilde{F}$
is defined in the full Fock space $\mathcal{H}$ and

\begin{eqnarray*}
Tr_{U(2)}(\widetilde{F})^{2} &=&(z\overline{z}+2\theta )^{-\frac{1}{2}}(%
\overline{z}_{2}F^{11}p_{00}F^{11}z_{2}+\overline{z}%
_{1}F^{11}p_{00}F^{11}z_{1})(z\overline{z}+2\theta )^{-\frac{1}{2}} \\
&=&-2(4\theta )^{2}(z\overline{z}+2\theta )^{-2}(z\overline{z}+4\theta
)^{-1}(z\overline{z}+6\theta )^{-1}.
\end{eqnarray*}

Inserting this expression into (\ref{kinstftild}) and using (\ref{REPN}), we
get the instanton number as a converge series of sum one.

Now we are ready to calculate explicitly the $U(2)$ instanton number (\ref%
{kinstftild}). First we observe from (\ref{Aexplicite}), (\ref{UCDU}) and (%
\ref{Fexplicite}) that the gauge field $\widetilde{A}$ and the field
strength $\widetilde{F}$ are well defined in the full Fock space\textit{\ }$%
P_{J}\mathcal{H\otimes H}$ and $Tr_{U(2)}(\widetilde{F})^{2}$ vanishes
rapidly enough at infinity (as $(z\overline{z})^{-4}$ in the region $n_{1},$ 
$n_{2}\longrightarrow \infty $ or $J\rightarrow \infty $ of the Fock space).
Then (\ref{kinstftild}) is represented by a converge series. These
properties permit us to use, as for ordinary Yang-Mills theory, the cyclic
property of the trace and the Leibniz rule of the exterior derivative to
rewrite (\ref{kinstftild}) under the form

\begin{equation}
k=-\frac{(4\pi \theta )^{2}}{8\pi ^{2}}Tr(\widetilde{F})^{2}=-\frac{(4\pi
\theta )^{2}}{8\pi ^{2}}Tr\left[ dK\right]  \label{SURFT}
\end{equation}%
where $K=\widetilde{A}\widetilde{F}-\frac{1}{3}(\widetilde{A})^{2}.$

Now, let $K=K_{\overline{z}_{1}}dz_{1}dz_{2}d\overline{z}_{2}+K_{\overline{z}%
_{2}}dz_{2}dz_{1}d\overline{z}_{1}+K_{z_{1}}d\overline{z}_{1}dz_{2}d%
\overline{z}_{2}+K_{z_{2}}d\overline{z}_{2}dz_{1}d\overline{z}_{1}$ be a
three form. The differential of $K$ is given by:

\begin{equation*}
dK=\frac{1}{2\theta }([\overline{z}_{\alpha },K_{z_{\alpha }}]+[z_{\alpha
},K_{\overline{z}_{\alpha }}])dz_{1}d\overline{z}_{1}dz_{2}d\overline{z}_{2}
\end{equation*}%
and its integration over a finite volume is expressed in the noncommutative
case by a trace over a finite Fock sub-space $\mathcal{H}_{V}\subset $ $%
\mathcal{H}$. Let $\mathcal{H}_{V}$ be a sub-space delimited by the
boundaries $P_{J_{2}}\mathcal{H}$ and $P_{J_{1}}\mathcal{H}$ with quantum
number $J_{2}\rangle J_{1}$. The Integration of $dK$ is given by

\begin{eqnarray}
Tr_{\mathcal{H}_{V}}dK &=&\frac{1}{2\theta }\sum_{J=J_{1}}^{J_{2}}%
\sum_{m=-J}^{J}\left\langle J,m\right\vert [\overline{z}_{\alpha
},K_{z_{\alpha }}]+[z_{\alpha },K_{\overline{z}_{\alpha }}]\left\vert
J,m\right\rangle  \notag \\
&=&\frac{1}{2\theta }\sum_{J=J_{1}}^{J_{2}}Tr_{P_{J}\mathcal{H}}([\overline{z%
}_{\alpha },K_{z_{\alpha }}]+[z_{\alpha },K_{\overline{z}_{\alpha }}]).
\label{TRHV}
\end{eqnarray}

By using (\ref{REPJM}) one can see that the terms coming from $Tr_{P_{J}%
\mathcal{H}}(\overline{z}_{\alpha },K_{z_{\alpha }})$ and $Tr_{P_{J}\mathcal{%
H}}(K_{\overline{z}_{\alpha }}z_{\alpha })$ cancel the terms coming from $%
Tr_{P_{J+\frac{1}{2}}\mathcal{H}}(K_{z_{\alpha }}\overline{z}_{\alpha })$
and $Tr_{P_{J+\frac{1}{2}}\mathcal{H}}(z_{\alpha },K_{\overline{z}_{\alpha
}})$ respectively, so that the contributions corresponding to the interior $%
J_{2}\rangle J\rangle J_{1}$ will be cancelled out to keep only
contributions coming from boundaries as

\begin{eqnarray}
Tr_{\mathcal{H}_{V}}dK &=&\frac{1}{2\theta }\sum_{m=-J_{2}}^{J_{2}}(\left%
\langle J_{2},m\right\vert \overline{z}_{\alpha }K_{z_{\alpha }}\left\vert
J_{2},m\right\rangle -\left\langle J_{2},m\right\vert K_{\overline{z}%
_{\alpha }}z_{\alpha }\left\vert J_{2},m\right\rangle )  \notag \\
&&-\frac{1}{2\theta }\sum_{m=-J_{1}}^{J_{1}}(\left\langle J_{1},m\right\vert
K_{z_{\alpha }}\overline{z}_{\alpha }\left\vert J_{1},m\right\rangle
-\left\langle J_{1},m\right\vert z_{\alpha }K_{\overline{z}_{\alpha
}}\left\vert J_{1},m\right\rangle .  \label{TRHV1}
\end{eqnarray}

This result is the noncommutative version of the Stokes' theorem. For the
trace over the full Fock space $\mathcal{H}$ $J_{1}=0$ and $%
J_{2}\longrightarrow \infty .$ Since $\overline{z}_{\alpha }\left\vert
0,0\right\rangle =0$ and $\left\langle 0,0\right\vert z_{\alpha }=0$, (\ref%
{TRHV1}) reduces to

\begin{eqnarray}
Tr_{\mathcal{H}}dK &=&\underset{J_{2}\longrightarrow \infty }{\lim }\text{ }%
\frac{1}{2\theta }\sum_{m=-J_{2}}^{J_{2}}(\left\langle J_{2},m\right\vert 
\overline{z}_{\alpha }K_{z_{\alpha }}\left\vert J_{2},m\right\rangle
-\left\langle J_{2},m\right\vert K_{\overline{z}_{\alpha }}z_{\alpha
}\left\vert J_{2},m\right\rangle )  \notag \\
&&\underset{J_{2}\longrightarrow \infty }{\lim }\frac{1}{2\theta }%
Tr_{P_{J_{2}}\mathcal{H}}(\overline{z}_{\alpha },K_{z_{\alpha }}-K_{%
\overline{z}_{\alpha }}z_{\alpha })  \label{ASYMPTR}
\end{eqnarray}%
from which we deduce the instanton number

\begin{equation}
k=-\frac{(4\pi \theta )^{2}}{8\pi ^{2}}Tr\left[ dK\right] =-\frac{(4\pi
\theta )^{2}}{8\pi ^{2}}\underset{J_{2}\longrightarrow \infty }{\lim }\frac{1%
}{2\theta }(Tr_{P_{J}}\otimes Tr_{P_{J_{2}}\mathcal{H}})(\overline{z}%
_{\alpha },K_{z_{\alpha }}-K_{\overline{z}_{\alpha }}z_{\alpha })
\label{kASYMP}
\end{equation}%
where $K_{z_{\alpha }}$ and $K_{\overline{z}_{\alpha }}$ are the components
of the three form $K=Tr_{U(2)}(\widetilde{A}\widetilde{F}-\frac{1}{3}(%
\widetilde{A})^{3})$. The relation (\ref{kASYMP}) shows that the calculus of
the instanton number is determined by the behavior of the components of the
three form $Tr_{U(2)}(\widetilde{A}\widetilde{F}-\frac{1}{3}(\widetilde{A}%
)^{3})$ in the asymptotic region $J\longrightarrow \infty $ of the Fock
space.

First (\ref{UCDU}) behaves like

\begin{eqnarray}
\underset{J\rightarrow \infty }{\lim }U^{\dagger }dU &=&-\frac{1}{2z%
\overline{z}}%
\begin{pmatrix}
\overline{z}_{1} & -2\overline{z}_{2} \\ 
0 & -\overline{z}_{1}%
\end{pmatrix}%
dz_{1}-\frac{1}{2z\overline{z}}%
\begin{pmatrix}
-\overline{z}_{2} & 0 \\ 
-2\overline{z}_{1} & \overline{z}_{2}%
\end{pmatrix}%
dz_{2}  \notag \\
&&+\frac{1}{2z\overline{z}}%
\begin{pmatrix}
z_{1} & 0 \\ 
-2z_{2} & -z_{1}%
\end{pmatrix}%
d\overline{z}_{1}+\frac{1}{2z\overline{z}}%
\begin{pmatrix}
-z_{2} & -2z_{1} \\ 
0 & z_{2}%
\end{pmatrix}%
d\overline{z}_{2}  \label{ASYMUDU}
\end{eqnarray}%
and has the same form as for the commutative case $g^{-1}dg$ where $g$ is
given by (\ref{USU2}). In this asymptotic region of the Fock space one may
see from (\ref{Fexplicite}) and (\ref{Aexplicite}) that the components of $%
F^{ab}$ and $\widetilde{F}^{ab}=(U^{\dagger }FU)^{ab}$ behave like $(z%
\overline{z})^{-2}$ and the components of the gauge fields $A^{ab}$ and $%
(U^{\dagger }AU)^{ab}$ behave like $(z\overline{z})^{-2}\overline{z}_{\alpha
}$ or $(z\overline{z})^{-2}z_{\alpha }$. Then from (\ref{Atilde}), the
asymptotic behavior of $U^{\dagger }AU$ and (\ref{ASYMUDU}) one deduces that
the gauge field $\widetilde{A}$ reduces to a pure gauge

\begin{equation*}
\underset{J\longrightarrow \infty }{\lim }\text{ \ }\widetilde{A}%
\longrightarrow U^{\dagger }dU
\end{equation*}%
leading to

\begin{equation}
\text{ }\underset{J\longrightarrow \infty }{\lim }Tr_{U(2)}K=\underset{%
J\longrightarrow \infty }{\lim }Tr_{U(2)}(\widetilde{A}\widetilde{F}-\frac{1%
}{3}(\widetilde{A})^{3})\longrightarrow -\frac{1}{3}Tr_{U(2)}((U^{\dagger
}dU)^{3})  \label{AAA}
\end{equation}%
where $U^{\dagger }dU$ is given by (\ref{ASYMUDU}). A straightforward
computation gives

\begin{eqnarray}
\underset{J\longrightarrow \infty }{\lim }Tr_{U(2)}K &=&+\frac{1}{(z%
\overline{z})^{2}}\overline{z}_{1}dz_{1}dz_{2}d\overline{z}_{2}+\frac{1}{(z%
\overline{z})^{2}}\overline{z}_{2}dz_{2}dz_{1}d\overline{z}_{1}  \notag \\
&&-\frac{1}{(z\overline{z})^{2}}z_{1}d\overline{z}_{1}dz_{2}d\overline{z}%
_{2}-\frac{1}{(z\overline{z})^{2}}z_{2}d\overline{z}_{2}dz_{1}d\overline{z}%
_{1}.  \label{TRUK}
\end{eqnarray}

Hence

\begin{equation}
K_{z_{\alpha }}=-z_{\alpha }\frac{1}{(z\overline{z})^{2}}\text{ , \ }K_{%
\overline{z}_{\alpha }}=\frac{1}{(z\overline{z})^{2}}\overline{z}_{\alpha
}\Longrightarrow \overline{z}_{\alpha }K_{z_{\alpha }}-K_{\overline{z}%
_{\alpha }}z_{\alpha }=-\frac{2}{z\overline{z}}\text{.}  \label{ZKZK}
\end{equation}

Inserting (\ref{ZKZK}) into (\ref{kASYMP}), we get%
\begin{eqnarray}
k &=&(Tr_{P_{J}\mathcal{H}}\otimes -\frac{(4\pi \theta )^{2}}{8\pi ^{2}}%
\frac{1}{2\theta }\underset{J_{2}\longrightarrow \infty }{\lim }Tr_{P_{J_{2}}%
\mathcal{H}})(\frac{-2}{z\overline{z}})  \notag \\
&=&(2J+1)\frac{(4\pi \theta )^{2}}{8\pi ^{2}}\frac{1}{2\theta }\underset{%
J_{2}\longrightarrow \infty }{\lim }\frac{2(2J_{2}+1)}{4\theta J_{2}}=(2J+1)
\label{KFINAL}
\end{eqnarray}%
where the first factor of the right side of (\ref{KFINAL}) represents the
dimension of the space $P_{J}\mathcal{H}$ which has been identified with the
vector space $%
\mathbb{C}
^{k}$, $k=2J+1$, in which the linear operators $B_{1}$ and $B_{2}$ of the
ADHM construction act. This number, $k=Dim(%
\mathbb{C}
^{k})=2J+1$, is considered as the instanton number of the ADHM construction.
The second factor may be viewed as the non commutative version of the
winding number. In fact one may see from (\ref{kASYMP}), (\ref{AAA}) and (%
\ref{USU2}) that the instanton number resembles the element of the third
homotopy group $\pi _{3}(SU(2))\approxeq 
\mathbb{Z}
$ . It is the noncommutative version of the winding number measured by the
surface integral at infinity.

\begin{equation}
k=\frac{1}{24\pi ^{2}}\int_{S_{\infty }^{3}}dS_{\mu }tr(g^{-1}\partial _{\nu
}g)(g^{-1}\partial _{\rho }g)(g^{-1}\partial _{\sigma }g)\varepsilon ^{\mu
\nu \rho \sigma }  \label{kclass}
\end{equation}%
where $g$ $=U$ belongs to the classical $id_{k}\otimes SU(2)$ group (\ref%
{USU2}), $g^{-1}dg$ is given by (\ref{ASYMUDU}) where $z_{\alpha }$ and $%
\overline{z}_{\alpha }$ are taken as c-number and $\partial R^{4}=S_{\infty
}^{3}$.

Notice that since the second factor of the right hand side of (\ref{KFINAL})
characterizes the noncommutative winding number $n=1$, the instanton number $%
k$ (\ref{KFINAL}) may be interpreted as a sum of $k$ $U(2)$ instantons of
noncommutative winding number $n=1$. Each term of this sum can be
calculated, with the same formalism presented in this section, by replacing $%
P_{J}$ (\ref{PROJM}) by the projector of rank one

\begin{equation}
P_{Jm}=\left\vert J,m\right\rangle \left\langle J,m\right\vert .
\label{P1JM}
\end{equation}

Much more, we can generalize (\ref{KFINAL}) for a winding number $n$ by
replacing, in this section, the partial isometry $U$ by $U_{n}=(U)^{n}$
which keeps the same partial isometry property (\ref{UUCROIX}), the
relations (\ref{UPO}) and the normalization (\ref{NORMPSIT}) for the
solution $\widetilde{\Psi }=\Psi U_{n}$ giving the gauge field $\widetilde{A}%
_{n}=\widetilde{\Psi }_{n}^{\dagger }d\widetilde{\Psi }_{n}$ and the field
strength $\widetilde{F}_{n}=d\widetilde{A}_{n}+(\widetilde{A}_{n})^{2}$
satisfying the relations

\begin{equation*}
\widetilde{A}_{n}=U_{n}^{\dagger }AU_{n}+U_{n}^{\dagger }dU_{n}\text{ and }%
\widetilde{F}_{n}=U_{n}^{\dagger }FU_{n}
\end{equation*}%
where the components of $A$ and $F$ are given by (\ref{Aexplicite}) and (\ref%
{Fexplicite}) respectively. For the same reasons presented above, in the
asymptotic region $J\rightarrow \infty $ the components of $U_{n}^{\dagger
}AU_{n}$ behave as those of $U^{\dagger }AU$ (like $z_{\alpha }(z\overline{z}%
)^{-2}$ or $\overline{z}_{\alpha }(z\overline{z})^{-2}$) and the components
of $\widetilde{F}_{n}$ behave as those of $\widetilde{F}$ (like $(z\overline{%
z})^{-2}$). This asymptotic behavior implies that $Tr(\widetilde{F}_{n})^{2}$
is given by a converge series and therefore we can apply the same process
leading to (\ref{kASYMP}) by replacing the asymptotic behavior of the three
form $K$ by

\begin{equation}
\text{ }\underset{J\longrightarrow \infty }{\lim }Tr_{U(2)}K_{n}=\underset{%
J\longrightarrow \infty }{\lim }Tr_{U(2)}(\widetilde{A}_{n}\widetilde{F}_{n}-%
\frac{1}{3}(\widetilde{A}_{n})^{3})\longrightarrow -\frac{1}{3}%
Tr_{U(2)}((U_{n}^{\dagger }dU_{n})^{3}).  \label{TRKn}
\end{equation}

Now let us show by induction that (\ref{TRKn}) leads to the value $n$ of the
winding number. This is certainly true for $n=1$ (\ref{KFINAL}). Suppose (%
\ref{TRKn}) leads to the winding number $n$ and establish it for $n+1$.

Let $A_{n}^{u}=U_{n}^{\dagger }dU_{n}$, then $A_{n+1}^{u}=U^{\dagger
}A_{n}^{u}U+U^{\dagger }dU$. In the asymptotic region $J\rightarrow \infty $
of the Fock space, $U$ behave like the classical group (\ref{USU2})
(c-number). Then we can use the cyclic properties of the trace and the
Leibniz rules of the exterior derivative to show that

\begin{equation}
\underset{J\longrightarrow \infty }{\lim }Tr_{U(2)}(A_{n+1}^{u})^{3}=%
\underset{J\longrightarrow \infty }{\lim }Tr_{U(2)}((A_{n}^{u})^{3}+(U^{%
\dagger }dU)^{3}-3d(U_{n+1}^{\dagger }dU_{n}dU)).  \label{TRUn3}
\end{equation}

Since the third term of the right hand side of (\ref{TRUn3}) is a total
derivative, it does not contribute to (\ref{kASYMP}). In fact, for a total
derivative three form $K$ i.e. $K=dH$ where $H$ is any two form , the
process of cancellation which has led to the boundary terms (\ref{ASYMPTR})
occurs in evaluating $Tr_{P_{J_{2}}\mathcal{H}}(\overline{z}_{\alpha
},K_{z_{\alpha }}-K_{\overline{z}_{\alpha }}z_{\alpha })$ which vanishes.
This is the noncommutative version of Stokes'theorem for a three dimensional
manifold without boundary. And therefore, from (\ref{TRUn3}) we deduce that
the winding number calculated from the left hand side of (\ref{TRUn3}) is $%
n+1$, a sum of the contribution coming from the first term of the right hand
side of (\ref{TRUn3}) which is supposed to be $n$ and the contribution
coming from the second term $\underset{J\longrightarrow \infty }{\lim }%
Tr_{U(2)}(U^{\dagger }dU)^{3}$ which give $n=1$ (\ref{KFINAL}).

Hence we can conclude that:

- in this formalism the $U(2)$ instanton number of noncommutative ADHM
construction is $kn$, the product of the dimension of the Fock sub-space $%
P_{J}\mathcal{H}$, $k=2J+1$, times the winding number $n$. It also can be
viewed as $k$ $U(2)-$instantons of winding number $n$ calculated by using
the projector (\ref{P1JM}).

- This result clarifies the geometrical picture of the noncommutative ADHM
instanton number and shows way it is the same value as commutative instanton
number.

-The $U(2)$ instanton number depends on the rank of projectors not on their
form. The projectors

\begin{equation*}
P=\sum_{n_{1}=N_{1}}^{N_{2}}\left\vert n_{1}+k,n_{1}+l\right\rangle
\left\langle n_{1}+k,n_{1}+l\right\vert \text{ \ }or\text{ \ }%
P_{J}=\sum_{m=M_{1}}^{m=M_{2}}\left\vert J,m\right\rangle \left\langle
J,m\right\vert
\end{equation*}

$\forall k,l\in 
\mathbb{N}
\geq 0$ and $\forall J=\frac{1}{2}(n_{1}+n_{2})=0,\frac{1}{2},1,...\infty \,$%
and $m=\frac{1}{2}(n_{1}-n_{2})$\ runs by integer steps over the range $%
-J\leq M_{1}\leq m\leq M_{2}\leq J$ and $M_{2}-M_{1}=k\leq J$, give the same
instanton number $k=N_{2}-N_{1}+1=M_{2}-M_{1}+1$ modulo the winding number $%
n $. This property resembles the noncommutative analogue of the invariance
under translations of the instanton positions.

\textbf{Acknowledgement}

I would like to thank M. Dubois-Violette for helpful discussion.

\end{document}